\documentclass[preprint,12pt]{elsarticle}

\usepackage{epsfig}
\usepackage{epstopdf}
\usepackage[T1]{fontenc}
\usepackage{amssymb}
\usepackage{amsmath}
\usepackage{amsfonts}
\usepackage{verbatim}
\usepackage{graphicx}
\usepackage{hyperref}
\usepackage{algorithmic}
\usepackage{algorithm}
\usepackage{array}
\usepackage{multirow}
\journal{ICCN-2013/ICDMW-2013/ICISP-2013}
\begin{document}
%
\title{Fault Detection for RC4 Algorithm and its Implementation on FPGA Platform 
 }

\begin{frontmatter}
\author{Rourab Paul\corref{cor1}\fnref{fn1}}

\author{Amlan Chakrabarti\fnref{fn2}}
 \author{Ranjan Ghosh\fnref{fn3}}
                                                     
\cortext[cor1]{Corresponding author}                             
\fntext[fn1]{rourabpaul@gmail.com, A.K.C. School of I.T., University of Calcutta, Kolkata.}
\fntext[fn2]{Senior Memeber IEEE, A.K.C. School of I.T., University of Calcutta, Kolkata.}
\fntext[fn3]{Dumkal Institute of Engineering and Technology,Basantapur Education Society, Murshidabad after retirement from the Institute of Radio Physics and Electronics, University of Calcutta, Kolkata}

\address{}

\begin{abstract}
In hardware implementation of a cryptographic algorithm, one may achieve leakage of secret information by creating scopes to introduce controlled faulty bit(s) even though the algorithm is mathematically a secured one. The technique is very effective in respect of crypto processors embedded in smart cards. In this paper few fault detecting architectures for RC4 algorithm are designed and implemented on Virtex5(ML505, LX110t) FPGA board. The results indicate that the proposed architectures can handle most of the faults without loss of throughput consuming marginally additional hardware and power. 
\end{abstract}
\end{frontmatter}

\section{Endomorphism}
RC4 algorithm is very simple and is widely used as a stream cipher. Today RC4 is a part of many network protocols, e.g. SSL, TLS, WEP, WPA and many others.There were many cryptanalysis to look into its key weaknesses \cite{DBLP:spaul} followed by many new stream ciphers \cite{t:good}. RC4 is still the popular stream cipher since it is executed fast and provides high security. It is believed that mathematically secure crypto algorithm becomes vulnerable while implementing it in hardware \cite{DBLP:journals/tcas/GhoshMC11}, since it becomes possible to extract secret information by introducing faults in a controlled fashion due to which on fault detection techniques turn out to be a key issue related to hardware implementation. Moreover, shrinking dimension of raw devices  induces Single Event Upset (soft error) which is termed as a change of logic state caused by ions or electro-magnetic radiation striking the device. The dense devices use more hardware components for faster processing and in turn cause increase of ion beam radiation as internal faults. This ion beam radiation causes state change in CLB\cite{mac:nic}. Usually two types of faults tolerant circuits are in use,one is Hardware Based Fault Tolerant (HBFT) circuits \cite{A:FPGA}\cite{B:FPGA}\cite{C:FPGA} and the other is Algorithm Based Fault Tolerant (ABFT) circuits \cite{D:FPGA}\cite{E:FPGA}\cite{F:FPGA}\cite{G:FPGA}.  For HBFT, faults are detected either at the CLB level or at the LUT level.  

A hamming code based fault detecting and correcting scheme is proposed for stream cipher like A5/1(GSM), E0 (Blue-tooth), RC4 (WEP), and W7 on hardware platform in article \cite{I11:FPGA}. It is not necessary that faults are always sourced from the system itself.
ABFT circuits at the communication level with specific reference to RC4 are proposed in  \cite{Ali05ft-rc4:a} and \cite{G:FPGA}. A sequence number padded to each cipher character of RC4 is proposed in  \cite{Ali05ft-rc4:a}. In\cite{G:FPGA} a method is proposed where data are stored in a matrix and 1 byte checksum is added to each of the rows and columns of the matrix. For multiple error detection they used Knight Checksum. Both the fault methods detect the fault after execution of the cipher text and thus take some additional time.\\
There exists quite a few literatures on AES Fault tolerance Scheme \cite{D:FPGA}\cite{E:FPGA}. The article \cite{D:FPGA} has tried to find out the contagious sections of the AES algorithm from which section the probability of error spreading is maximum. It has been observed how a single bit and multiple bit errors can spread over the data with algorithm iterations.  Fault is located followed by its detection. They have introduced a parity checker scheme (16 bits) with input data block (16 bytes) which can detect single bit errors and as well as odd multiple bit errors.The error detecting efficiency of this scheme is not so good for efficient error detecting crypto system. For key scheduling process \cite{D:FPGA} has proposed an inverse key scheduling module which error detecting efficiency is appreciable but resource usage becomes twice of the original key scheduling model. 
In \cite{E:FPGA} three types of fault detection scheme based on cycle redundancy checks (CRC) are proposed. \\
\begin{figure}[]
\centering
\vspace{-2pt}
\includegraphics[width=12cm,height=5cm]{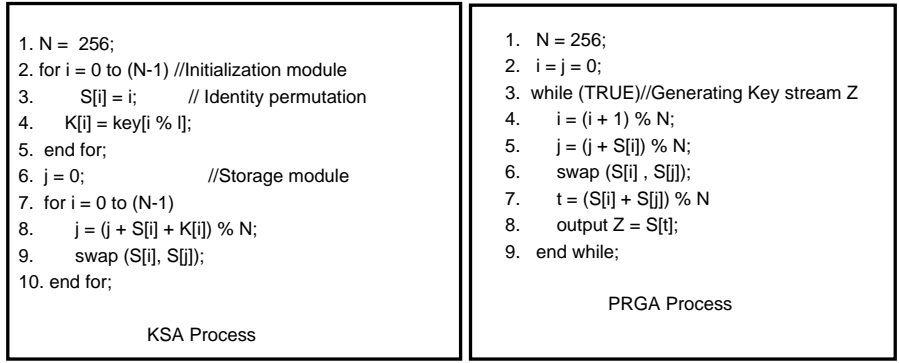}
\vspace{-2pt}
\caption{RC4 Algorithm}
\vspace{-2pt}
\label{fig:ksa_prga_fig}
\end{figure}
In this paper an ABFT scheme is proposed for RC4 stream cipher in which efficient fault detecting additional hardware blocks are designed with an intention to detect maximum errors using minimum resources.  The proposed scheme can detect faults at the very instant the ciphering is being executed. Fault blocks and the algorithm block are executed in parallel due to which the throughput remains unchanged.  When fault is detected, the system is reset to aware the user. Here faults are only detected,not corrected. In the absence of fault detecting blocks, occurrences of faults would cause changes in the power and timing parameters which would provide information to side channel attackers to extract information related to secret key. 
The paper is organized as follows: Section 2  details the overview of Fault techniques.The experimental results are summarized in Section 3. Conclusion and References are enlisted in Section 4 and 5.
\section{Fault detection techniques adopted for RC4 }
\begin{figure}[!htb]
\centering
\vspace{-2pt}
\includegraphics[width=8.5cm,height=5cm]{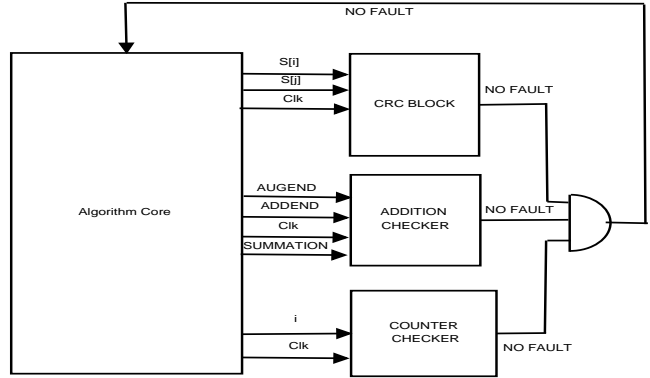}
\vspace{-2pt}
\caption{Fault Blocks}
\vspace{-2pt}
\label{fig:fault_block}
\end{figure}

The RC4 has two sequential algorithms, namely KSA (Key Scheduling Algorithm) and PRGA (Pseudo Random Generator Algorithm) and are shown in Figure \ref{fig:ksa_prga_fig}. In case of RC4 a single or multiple bit errors can change the value of $'j'$ (see line 8 of KSA and line 5 of PRGA of Figure \ref{fig:ksa_prga_fig}) randomly which can addresses a wrong S-box element for further swapping process. So there is no correlation between number of faulty bit and the number of iteration of the algorithm.
 As we see in  Figure \ref{fig:ksa_prga_fig}, RC4 has an identity S-Box S[N], N=0 to 255 and a secret key, key[l] where l is typically between 5 and 16, used to scramble the S-Box [N]. 
The purpose of the KSA-PRGA processes is to scramble the S-Box [N]. As both the processes have more or less identical operations, the design of fault tolerance modules is discussed for one process only.The detail hardware architecture of the core algorithm has been described else where in \cite{DBLP:journals_rourab}.\\
Arithmetic and logical operations are very much fault prone.The three common arithmetic operations of KSA and PRGA are $'i'$, $'j'$ computation and retrieval of $S[i], S[j]$ for swapping process. Any malfunction in these operations may cause wrong encryption/decryption results which may be a clue for an attacker. It has been seen that  $'i'$,depends on a plane binary up counting process, where as  $'j'$,  depends on an addition operation. Any abnormality in$'i'$, $'j'$ computation can address wrong $S[i] and S[j]$ resulting in wrong swapping process. Any single or multiple faults on S-Box can spread all across the algorithm quickly hampering the algorithm randomness as well as the cipher text authenticity.\\
To check the  $'i'$ functionality according to the algorithm,a new efficient counter checker module is proposed. For  $'j'$ computation an efficient addition checker is proposed.The correct computation of $S[i] and S[j]$ is ensured by using a CRC checker on S-box [N].The three checkers are executed parallel with the main algorithm without hampering the algorithm throughput. The proposed three fault blocks are shown in Figure \ref{fig:fault_block} and the structure of the CRC code is shown in Figure \ref{fig:crc_data}. Before the execution of each round, the core algorithm checks the “no fault”signal contributed by the three proposed fault checkers.  Each of the three fault checkers feed $“no\_fault”$ to the algorithm block through AND gate.Any fault detected by a particular fault module can stop the execution of the algorithm at that instant of clock edge.
\subsection{Error detection on S-Box: CRC Checker}
To detect fault on S-Box Array standard CRC technique of 4-degree polynomial is used. Lower degree of polynomial is used to reduce length of redundant CRC bits. It has been seen that CRC has good efficiency to detect single bit errors, double bit errors and odd number of errors. A dedicated hardware block to
 execute CRC algorithm has not been designed since that would require a huge hardware resource,large computation power and might cause some synchronization problem with the main algorithm.This synchronization problem is a sensitive issue as the main crypto core has a very high throughput based on dual edge sensitivity. To bye-pass, a standard 4-degree divisor, $X^4+X^3+1$ is chosen and four bit residue is computed as CRC code which has been padded to each S-Box element, each element thus becomes a 12-bit data instead of 8-bit, as shown in Figure \ref{fig:crc_data}. This new S-box is stored as two S-Boxes in the CRC hardware block (vide Figure \ref{fig:crc_block}) as well as in main algorithm S-Box. CRC block has two input S[i] and S[j]. In each clock S[i] and S[j] is computed by the main algorithm block and then it has been checked by CRC module whether S-Box element is correct or not. If there is no error in the CRC block it will proceed for the next clock cycle. According to [15] CRC checker can detect all odd number of errors since the divisor polynomial can be
\begin{table}[]
\caption{CRC} 
\centering  
\resizebox{11 cm}{!}{%

    \begin{tabular}{ c c c c  }
        \hline
      \#\ &  Possible combination       & \multicolumn{2}{c}{CRC}\\  
\cline{3-4}
 faulty bit(bit)& of faulty bit & Detected Fault   & Undetected Fault    \\ \hline  
              
1	& 8C1=8	& 8	&0\\
2	& 8C2=28	& 21	& 7\\
3	& 8C3=56	& 56	& 0\\
4	& 8C4=70	& 70	& 0\\
5	& 8C5=56	& 56	& 0\\
6	& 8C6=28	& 0	& 28\\
7	& 8C7=8	&8	& 0\\
8	& 8C8=1	& 0	& 1\\\hline
Total fault	& 255	&219	&36\\ \hline
Error Detecting Efficiency (\%)\ &~& 86\\
  \hline
       
    \end{tabular}
}
\label{table:crc} 
\end{table}
 divided by X+1. It can detect all isolated double bit errors, since the polynomial cannot be divided by$X^t+1$ (where t=2 to 8). Table \ref{table:crc} shows the number of detected faulty bit, maximum number of faulty bit combination and the number of detected and undetected fault using the CRC based method. As shown in table 1 the efficiency evaluates 86\%.
\subsubsection{Hardware design of CRC checker}
\begin{figure}[!htb]
\centering
\vspace{-2pt}
\includegraphics[width=9cm]{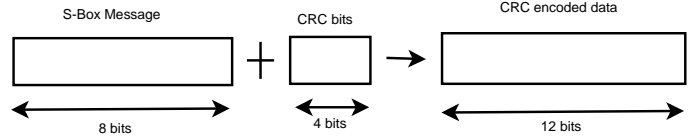}
\vspace{-6pt}
\caption{CRC encoded data}
\vspace{-2pt}
\label{fig:crc_data}
\end{figure}

\begin{figure}[!htb]
\centering
\vspace{-6pt}
\includegraphics[height=7cm,width=12cm]{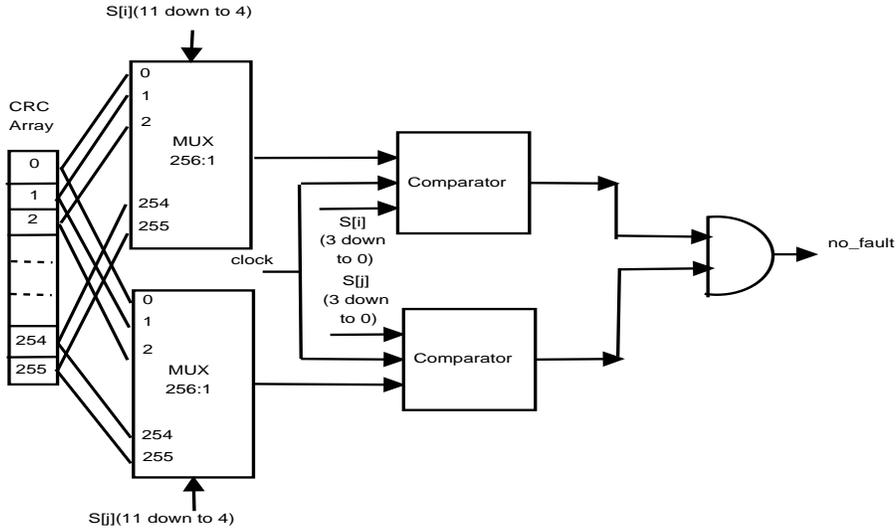}
\vspace{-6pt}
\caption{CRC hardware block}
\vspace{-6pt}
\label{fig:crc_block}
\end{figure}

The input of CRC block is $S[i] and S[j]$ which are of 12 bit, 8 bit is for data and 4 bit for CRC code. The CRC encoded data format is shown in \ref{fig:crc_data}. A look up table of 4 bit width is already stored in $'buffer'$ called CRC array with appropriate CRC code. The $S[i]$ and $S[j]$ port of
 algorithm block is connected with  $S[i]$ and $S[j]$ of CRC block. In every rising edge of clock the CRC encoded  $S[i]$ and $S[j]$ data from the algorithm block has been checked with CRC array by the CRC  block.The CRC hardware architecture is shown in \ref{fig:crc_block}.  

\subsection{Error Detection on Addition Checker} 
There are few standard types of error detecting techniques on arithmetic operations \cite{I13:FPGA} such as parity and residue techniques which are most popular due to its low cost and high error detecting efficiency. Residue technique is motivated on modulus operation which costs huge hardware footprint \cite{rourab:paul} in FPGA based platform. This is the reason that the parity scheme is adopted in this paper for addition checker. The 8-bit data is split into two 4-bit nibble to increase the error detecting efficiency as is seen in  table \ref{table:redundancy}.
\begin{table}[htb]
\caption{Degree of Redundancy of proposed error detecting scheme} 
\centering  
\begin{tabular}{c c c} 
\hline\hline                        
Type of EDC &Number of Redundant Bits & Degree of Redundancy \\ [0.5ex] 
\hline                  
Byte parity  & 1 & 1/8=12.5 percent   \\ 
Nibble Parity & 2 & 2/8=25 percent   \\      
\hline 
\end{tabular}
\label{table:redundancy} 
\end{table}
Now it is necessary to describe how the parity prediction scheme is initiated for addition operation. Of the two 8 bit numbers, such as If there are two numbers of 8 bit width, such as \textit{add} and \textit{aug}. 4 parity bit such as then \textit{p(add lower), p(add higher), p(aug lower)} and \textit{p(aug higher)} following manner. \\
\textit{p(add lower)=p(add(0) xor add(1) xor add(2) xor add(3)),}\\
\textit{p(add higher)=p(add(4) xor add(5) xor add(6) xor add(7)),}\\
\textit{p(aug lower)=p(aug(0) xor aug(1) xor aug(2) xor aug(3)),}\\
\textit{p(aug higher)=p(aug(4) xor aug(5) xor aug(6) xor aug(7)).}
\subsubsection{Prediction for arithmetic addition}It is well known that the parity of the sum of two natural number can be obtained by XORing the parities of both summands and of all carries propagated between any two adjacent bits, plus the possible carry-in into the least significant position. Hence\\

$p(add lower+ aug lower) = p(add lower) xor p(aug lower) xor Cin xor \bigoplus_{i=0}^{3}{C^{(i)}_{out}}\\
p(add higher+ aug higher) = p(add higher) xor p(aug higher) xor Cin xor \bigoplus_{i=4}^{7}
{C^{(i)}_{out}}$
\subsubsection{Hardware strategy of addition checker}
In KSA process the inputs of addition checker such as, $'j'$, $S[i]$, and $K[i]$ and its summation result has been passed to the addition checker block. By parity prediction technique the addition checker fault block can check the  whether the summation is right or wrong. The efficiency is about 75\%\ which is portrayed
\begin{table}[htb]
\caption{Addition Checker} 
\centering  
\resizebox{11 cm}{!}{%

    \begin{tabular}{ c c c c  }
        \hline
     \#\  &  Possible combination       & \multicolumn{2}{c}{Counter Checker}\\  
\cline{3-4}
 faulty bit(bit)& of faulty bit & Detected Fault   & Undetected Fault    \\ \hline               
1	& 8C1=8	& 8	&0\\
2	& 8C2=28	& 16	& 12\\
3	& 8C3=56	& 56	& 0\\
4	& 8C4=70	& 32	& 38\\
5	& 8C5=56	& 56	& 0\\
6	& 8C6=28	& 16	& 12\\
7	& 8C7=8	&8	& 0\\
8	& 8C8=1	& 0	& 1\\\hline
Total fault	& 255	&192	&63\\ \hline
Error Detecting Efficiency (\%)\ &~& 75\\
  \hline
       
    \end{tabular}
}
\label{table:Addition} 
\end{table}
 in table \ref{table:Addition}. The same addition checker module has been used for Z computation in line no. 5 and 7 of PRGA process.
\subsection{Error Detection on i counter} 
Several techniques\cite{I14:FPGA} have already been developed in order to improve the reliability of binary counter. A completely new technique is proposed consuming very low resource usage and exhibit very high error detecting efficiency. An interesting pattern has been observed in binary counting. If the parity of even bit position data is computed, the parity of  first 4 set of data will be the complement of next 4 set of data.Similarly if the parity of odd bit position data is computed, the parity of first 4 set of data will also be the complement to next 4set of data. There exist another pattern in parity bit of msb of upper nibble and msb of lower nibble. The parity pattern is shown in table \ref{table:pattern}. Following this pattern we have designed a fault checker on the $'i'$ counter which store the 8 consecutive counting of $'i'$ and feed a decision whether the counting is right or wrong based on the pattern prediction that we describe in table \ref{table:pattern}.
\begin{table}[]
\caption{binary counting pattern} 
\centering  
\resizebox{7 cm}{!}{%

    \begin{tabular}{ c c c c c }
        \hline
      counting number        &  parity      \\ \hline                     
             ~              & MSB of two nibble  & Even    & Odd     \\ \hline
       00000000                     & 0                    &0     & 0         \\ 
       00000001                     & 0                    & 1    & 0          \\   
       00000010                     & 0                    & 0    & 1            \\  
       00000011                     & 0                    & 1    & 1           \\ \hline
       00000100                     & 0                    & 1    & 1 \\
       00000101                    & 0                     & 0    & 1 \\
       00000110                    & 0                     & 1    & 0 \\
       00000111                    & 0                     & 0    & 0 \\ \hline
       00001000                    & 1                     &0     & 1 \\
       00001001                    & 1                     & 1    & 1 \\
       -----                             & --                    &--     & --\\
       11111110                    & 1                    & 1     & 0 \\
       11111111                    & 0                    & 0     & 0 \\
\hline      
    \end{tabular}
}
\label{table:pattern} 
\end{table}
\subsubsection{Hardware overview of counter checker}
The main RC4 algorithm core increasing $'i'$ in every clock cycle. Every 8 set of data has been buffered into an array in counter checker fault block. The fault block separate 8 set of data into two 4 set of part. The fault checking algorithm checks the proposed pattern mentioned in table \ref{table:pattern} after every 8 clock cycle and make decision that whether fault has been occurred or not which has been fed to the main algorithm block.The error detecting efficiency of proposed counter checker has been shown in table \ref{table:counter}.
\begin{table}[]
\caption{Counter Checker} 
\centering  
\resizebox{11 cm}{!}{%

    \begin{tabular}{ c c c c  }
        \hline
      \#\  &  Possible combination       & \multicolumn{2}{c}{Counter Checker}\\  
\cline{3-4}
 faulty bit(bit)& of faulty bit & Detected Fault   & Undetected Fault    \\ \hline  
              
1	& 8C1=8	& 8	&0\\
2	& 8C2=28	& 20	& 8\\
3	& 8C3=56	& 56	& 0\\
4	& 8C4=70	& 55	& 15\\
5	& 8C5=56	& 56	& 0\\
6	& 8C6=28	& 20	& 8\\
7	& 8C7=8	&8	& 0\\
8	& 8C8=1	& 1	& 0\\\hline
Total fault	& 255	&224	&31\\ \hline
Error Detecting Efficiency (\%)\ &~& 88\\
  \hline       
    \end{tabular}
}
\label{table:counter} 
\end{table}
\section{Results and discussion}
The individual fault blocks have been implemented on Xilinx Virtex5 FPGA board. The resource consumption of 3 fault blocks is very less compared to the main architecture. Of the proposed fault blocks, the counter checker block and add checker sub-blocks takes very less resource compared to main architecture (0.09\%\ \&\ 0.31\%\ ) while the main  CRC checker sub-block(50\%\ ) is resource hungry and takes considerably high resource compared to main architecture. Not only this, the 3 blocks have 45\%\, 0.2\%\ \&\ 0.26\%\ LUTs usage compare to the main architecture.
The detail estimation of resource usage is given in table \ref{table:resource}.

\begin{figure}[]
\centering
\vspace{-2pt}
\includegraphics[width=12cm,height=5cm]{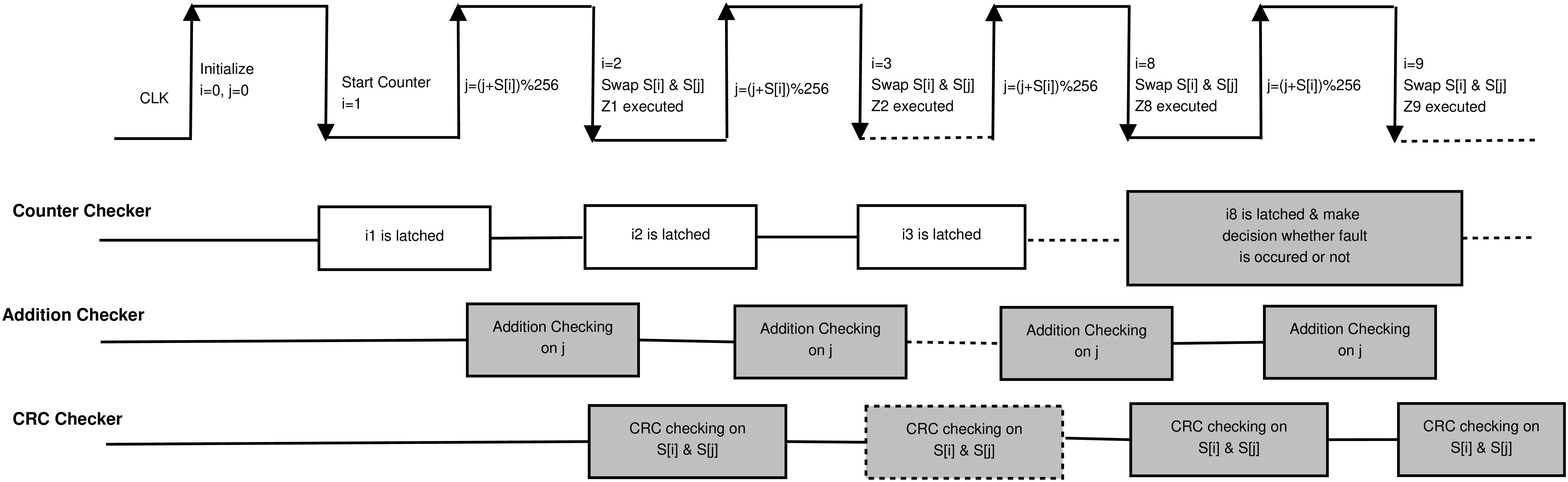}
\vspace{-6pt}
\caption{Timing diagram of proposed fault modules with respect to to main algorithm}
\vspace{-2pt}
\label{fig:timing_fig}
\end{figure}
The xilinx xpower tool to measure system power consumption \cite{xilinx:online1}. Three fault blocks, CRC Counter Checker \&\ Addition Checker consume 7\%\, 1.2\%\ \&\ 4.3\%\ power compared to main architecture power. Resource utilization table \ref{table:resource} \& power consumption table \ref{table:power} is showing that the additional fault blocks has very less resource utilization and less power consumption which is the desirable goal of such kind fault detection application on FPGA based platform.
\begin{table}[htb]
    \begin{minipage}{.51\linewidth}
      \caption{power consumption }
\resizebox{7 cm}{!}{%
      \centering
   \begin{tabular}{ |c|c|c|c|c|c|c|c|}
        \hline
      Power                                    & Main                 & CRC         & Counter        & Addition\\
      ( milli watt) &                           Core         &~             & ~                    &Checker \\
 \hline
        Total                                     & 994.72               & 70.58    &  12.81       & 43.25 \\  
        Power                                               &~                         &~           &~             &~  \\\hline
       Quiescent                                         & 914.87                & 30.71   & 0.67       & 2.29      \\  
       Power                                               &~                         &~           &~             &~  \\\hline
        Dynamic                                & 52.86                 & 67.1     & 11.9        & 40.96\\ 
       Power                                               &~                         &~           &~             &~  \\\hline
       Clock r                                      & 47.33                 & 19.03   & 7.11         &0.19      \\ 
       Power                                               &~                         &~           &~             &~  \\\hline
       Logic                                          & 0.60	          &4.47     &0.06          &0.13\\ 
       Power                                               &~                         &~           &~             &~  \\\hline
       IOs                                         &  0.60	           &4.47     &0.06         &0.13\\ 
        Power                                               &~                         &~           &~             &~  \\\hline
       Signal                                    &4.74	           &43.39     &4.94        &41.66\\  
              Power                                               &~                         &~           &~             &~  \\
        \hline
    \end{tabular}
}
\label{table:power}
    \end{minipage}%
    \begin{minipage}{.5\linewidth}
      \centering
        \caption{Resource utilization }
\resizebox{7 cm}{!}{%
      \begin{tabular}{ |c|c|c|c|c|}
        \hline
       Logic                                       & Main                 & CRC    & Counter     & Addition  \\ 
        Usage $\#\ $                              & Core      &~       &Checker            &Checker\\\hline
        Slice                                            & 4139                     &2042    & 4                & 13  \\  
        Register                                   &~                            &~         &~                 &~ \\ \hline
        slice                                         & 12560                       & 5653           & 26         & 33      \\   
        LUT                                              &~                            &~                   &~           &~ \\ \hline
       fully                                          & 4132                     & 2034              & 52              & 78\\  
      used LUT                                  &~                            &~         &~                 &~ \\
     -FF pairs                                 &~                            &~         &~                 &~ \\
        \hline
    \end{tabular}
}
\label{table:resource}
    \end{minipage} 
\end{table}
In an earlier paper \cite{DBLP:journals_rourab} the RC4 algorithm was implemented in hardware using Vertex 5 FPGA in which 1-byte in 1-clock was the approximate execution speed which has been achieved by carrying the addition process (line 5 of PRGA process) during the rising edge of a clock pulse and the swapping and key streams generation (lines 6 and 7 of PRGA process)during falling edge of the same clock pulse with a loss of one initial clock pulse.The timing diagrams of the proposed three fault modules with respect to to main algorithm clock  are also shown in Figure \ref{fig:timing_fig}. At falling edge of every 8th consecutive clocks,$'i'$ is checked and at every rising edge,the addition checker and at every falling edge, the CRC checker is executing their respective tasks.  It becomes evident that the fault modules are so designed in hardware here that the throughput of the main RC4 algorithm remains unchanged.
\section{Conclusion}
In this paper three low cost fault block are designed for RC4 and implemented in FPGA operating concurrently with the progress of  the main algorithm consuming low power and resources, providing run time fault detection efficiency without affecting its throughput. On detection of even one fault the algorithm ceases execution. Had the main algorithm and fault blocks are executed sequentially, the throughput would have been reduced and the attacker would have been able to get the secrets of the algorithm observing power and timing parameters.  
\section{References}
\bibliographystyle{IEEEtran}
\bibliography{IEEEexample}
\end{document}